\documentclass{aip-cp}

\usepackage[numbers]{natbib}
\usepackage{rotating}
\usepackage{graphicx}

\begin{document}

\title{Search for Ultra-High Energy Photons with the \\Pierre Auger Observatory}

\author[aff1]{Daniel Kuempel\corref{cor1}}
\author[aff2,aff3]{for the Pierre Auger Collaboration}

\affil[aff1]{RWTH Aachen University,\\III. Physikalisches Institut A Otto-Blumenthal-Str., 52056 Aachen, Germany}
\affil[aff2]{Observatorio Pierre Auger, Av.\ San Mart\'{i}n Norte 304, 5613 Malarg\"ue, Argentina}
\affil[aff3]{Full author list: http://www.auger.org/archive/authors\_2016\_07.html}
\corresp[cor1]{Email: kuempel@physik.rwth-aachen.de}

\maketitle

\begin{abstract}
The Pierre Auger Observatory, located in Argentina, provides an unprecedented integrated aperture for the search of photons with energy above 100 PeV. In this contribution recent results are presented including the diffuse search for photons and the directional search for photon point sources. The derived limits are of considerable astrophysical interest: Diffuse limits place severe constraints on top-down models and start to touch the predicted GZK photon flux range while directional limits can exclude the continuation of the electromagnetic flux from measured TeV sources with a significance of more than 5$\sigma$. Finally, prospects of neutral particle searches for the upcoming detector upgrade AugerPrime are highlighted.
\end{abstract}

\section{INTRODUCTION}
Photons are unique messengers for exploring the universe and have revealed many new insights of astrophysical processes. The currently observed maximum energy of an astrophysical photon is in the TeV regime ($1~{\rm TeV} = 10^{12}~{\rm eV}$) detected by ground based Cherenkov telescopes (e.g.\ \cite{Degrange2015587}). At ultra-high energies (UHE) only charged cosmic rays with energies above $\sim 1$~EeV ($1~{\rm EeV} = 10^{18}~{\rm eV}$) have been observed. These cosmic rays can interact with low energy background photons (e.g.\ from the cosmic microwave background radiation, the radio background or in dense photon fields in the vicinity of a source) and produce a flux of UHE photons that can propagate for several Mpc without being absorbed. The main production process at UHE are photo-pion interactions where a nucleon $N$ interacts with a background photon $\gamma$ producing charged and neutral pions $N + \gamma \longrightarrow N + \pi$ in a GZK-type process \cite{GreisenPhysRevLett,Zatsepin:1966jv}. In the channel that conserves the charge of the original nucleon mostly neutral pions are produced which decay into secondary gamma rays $\pi^0 \rightarrow \gamma + \gamma$. 

Other predictions of UHE photons arise from non-acceleration models where the primary process is given by the decay (or annihilation) of primordial relics such as topological defects \cite{HILL1983469,Hindmarsh0034}, super heavy dark matter \cite{PhysRevLett.79.4302,BIRKEL1998297,Kuzmin1998,Blasi200257}, $Z$-burst scenarios \cite{WeilerPhysRevLett,WEILER1999303,Fargon0004} or other top-down models \cite{PhysRevD.74.115003,FodorPhysRevLett,Sarkar2002495,Barbot20035,AloisioPhysRev}. The expected photon fluxes in conventional acceleration scenarios depend on the composition and maximum energy of the cosmic rays at the source, the emissivity distribution and cosmological evolution of the production sites. Thus, the observation of UHE photons or limits on photon fluxes can pose constraints on the origin of UHE cosmic rays and the properties of the source as well as on non-acceleration models.

The Pierre Auger Observatory \cite{ThePierreAugerCollaboration2015172}, located in western Argentina near Malarg\"ue and centered at latitude 35.2$^\circ$ S and longitude 69.5$^\circ$ W at an altitude of about 1400~m above sea level, provides an unprecedended sensitivity to search for photons at UHE due to its vast collecting area and its ability to discriminate between photon and hadron induced cosmic-ray air showers. It is a hybrid system, a combination of a large surface detector array (SD) and a fluorescence detector (FD). The SD consists of 1660 water Cherenkov particle detectors covering an area of 3000~km$^2$ on a triangular grid with 1.5~km spacing. The FD includes 27 individual florescence telescopes overlooking the area from five different sites. While the SD has a duty cycle of nearly 100\%, the FD can only operate on dark nights resulting in an duty cycle of 10-15\%. 

The paper is organized as follows: First, latest results on the search for photons are highlighted without taking into account the directional information of the event. These ``diffuse'' searches place limits on the fraction and flux of UHE photons and are of considerable interest testing GZK and non-acceleration models. Second, the search for photon point sources by taking the directional information of the event into account are presented. Photon flux limits from point sources can be used to constrain the continuation of measured photon spectra in the TeV regime to the EeV regime and place upper limits on the spectral cutoff energy. Finally, the results are summarized and a brief outlook on the prospects of photon searches for the planned upgrade AugerPrime is given. 

\section{DIFFUSE PHOTON SEARCHES}
Photon induced cosmic air showers are characterized by a lower average number of muons and an on average later shower development compared to hadron induced showers. The reduced number of muons can be explained by a two orders of magnitude lower radiation length compared to the mean free path for photo-nuclear interactions where energy is transferred into the hadron/muon channel. The delayed shower development is caused by the typically small multiplicity of electromagnetic interactions. 

Using dedicated detector specific FD and SD observables to discriminate between hadron and photon induced air showers upper limits on the photon fraction and fluxes could be set \cite{Abraham2007155,Abraham2008243,Abraham2009399,SettimoICRC}. Upper limits on the photon fraction at EeV energies are a few percent, e.g.\ 2\% above $10^{19}$~eV using SD data \cite{Abraham2008243} (at 95\% c.l.). Latest results on photon flux limits above 10 EeV using SD data \cite{BleveICRC} are illustrated in the following in more detail. Above this energy the Landau-Pomeranchuk-Migdal (LPM) effect becomes important \cite{Landau:1953um,Landau:1953gr,PhysRev.103.1811}. The basic principle is that the Bethe-Heitler cross-section $\sigma_{\rm BH}$ for pair production by photons can be reduced due to destructive interference from several scattering centers and the shower development is further delayed. Photons can also convert into $e^{\pm}$-pairs in the geomagnetic field of the Earth inducing a pre-shower \cite{PhysRevD.24.2536}. The probability is site dependent and the threshold energy is $\sim 50$~EeV for the Pierre Auger Observatory \cite{Homola2007174}. Both effects are taken into account in Monte Carlo simulations of photon air showers used in this analysis. To ensure a good energy and profile reconstruction of the events several quality cuts are defined, e.g.\ only events in the zenith range between $30^\circ < \theta < 60^\circ$ and a maximum depth of shower maximum $X_{\rm max}$ not to be more than 50~g~cm$^{-2}$ below the ground level are taken into account. A full set of cuts is given in \cite{BleveICRC}. The energy of the event is related to the observable $S(1000)$, the signal in VEM (vertical equivalent muons) at 1000~m from the shower axis \cite{Newton2007414}. However, the conversion from ($S(1000),\theta$) to energy is specific for photon induced showers and an iterative procedure, similar to the one in \cite{Billoir:2007vq} is applied. 

\begin{figure}[ht!]
  \centerline{\includegraphics[width=460pt]{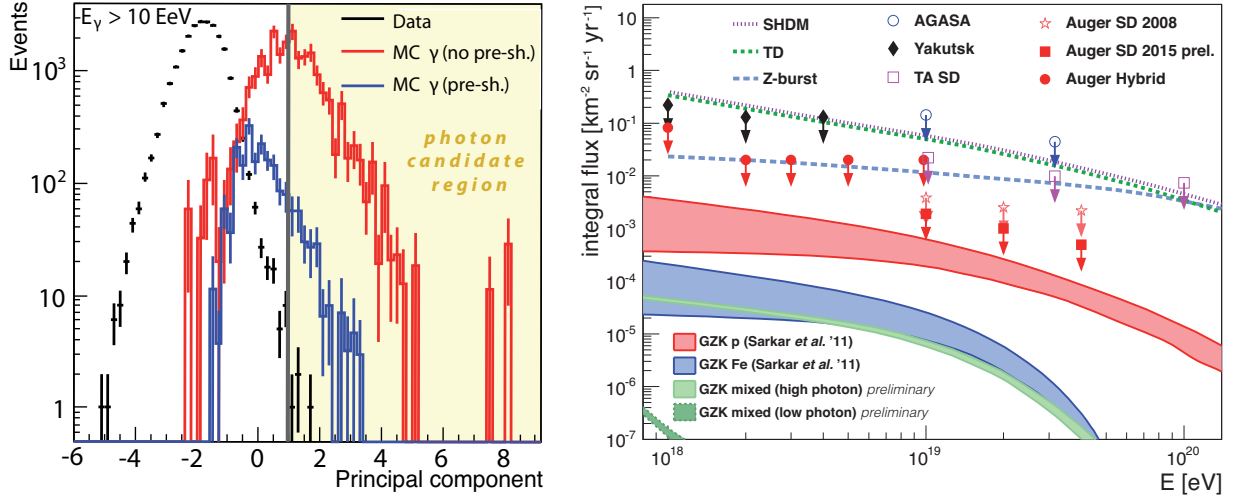}}
  \caption{\textit{Left panel:} Distribution of the principal component axis for photon energies $> 10$~EeV. The blue and red histograms illustrate photon Monte Carlo simulations with and without pre-showered events. The back histogram is the search data sample (98\% of the total, i.e.\ 22853 events). The \textit{a priori} cut value for photon candidate events is shown by a vertical line. (modified from \cite{BleveICRC}) \textit{Right panel:} Integral upper limits of the photon flux of AGASA \cite{PhotonAGASA}, Yakutsk \cite{PhysRevD.82.041101},TA \cite{PhysRevD.88.112005}, and Auger \cite{Abraham2008243,SettimoICRC,BleveICRC}. Predictions for non-acceleration models are given for SHDM \cite{PhysRevD.74.115003}, TD \cite{PhysRevD.74.115003}, and $Z$-burst \cite{Gelmini2008}. GZK photon predictions are shown as solid area, a pure proton and iron \cite{Kampert2012660,SarkarICRC} and two mixed scenarios, see text for details.}
  \label{fig.Diffuse}
\end{figure}

To select photon candidate showers, two event observables are defined, sensitive to the specific characteristics of purely electromagnetic showers. One observable measures the departure from the average lateral distribution function (LDF). At large distances from the shower axis photon induces showers have typically less signal compared to hadron induced showers of the same energy. The observable is defined as 
\begin{equation}
L_{\rm LDF} = \log_{10} \left( \frac{1}{N} \sum_i \frac{S_i}{{\rm LDF}(r_i)} \right)~~,
\end{equation}
where $i$ runs over surface detector stations $i = 1, ... , N$ with a radial distance $r_i > 1000$~m from the core, $S_i$ is the total signal of the $i^{\rm th}$ station and LDF$(r_i)$ the signal at distance $r_i$ according to the LDF fit. The second observable is defined as the risetime, i.e.\ the time difference between the 50\% and 10\% time quantiles of the FADC time trace. For geometrical reasons this parameter is sensitive to the deeper $X_{\rm max}$ of photon induced showers (larger risetime) and further increased by the larger contribution of the electromagnetic component. This observable needs to be corrected by azimuthal asymmetry effects obtaining $t_{1/2}$. The final observable is
\begin{equation}
\Delta = \left( \sum_i \frac{\delta_i}{N} \right)~~{\rm with}~~\delta_i = \frac{t_{1/2}-t^{\rm bench}_{1/2}}{\sigma_{t_{1/2}}}~~.
\end{equation}
Here, $t^{\rm bench}_{1/2}$ and $\sigma_{t_{1/2}}$ is the average corrected risetime and sampling fluctuations for data. Additional quality criteria are required and explained in detail in \cite{BleveICRC}. These two observables are combined in a principal component analysis (PCA) where 2\% of the data are used for training the classifier. The distribution of data and dedicated photon simulations weighted to an $E^{-2}$ power spectrum as a function of the principal component axis is shown in Figure \ref{fig.Diffuse} (left). Photon candidate events are \textit{a priori} defined as to be larger than the median of the weighted distribution of non-pre-showering photon simulations, cf.\ Figure \ref{fig.Diffuse} (left). 

As can be seen in Figure \ref{fig.Diffuse} (left), four events fulfill the photon candidate criteria using a data period between January 1 2004 and May 15 2013. Limits on the integral photon flux can be set by calculating $F_\gamma^{\rm CL} (E_\gamma > E_0) = N_\gamma^{\rm CL} / \langle \mathcal{E} \rangle$, where $N_\gamma^{\rm CL}$ is the Feldmann-Cousins upper limit of the number of photon events at a confidence level CL and $\langle \mathcal{E} \rangle$ the spectrum weighted average exposure derived by application of the same criteria to simulated showers (assuming a spectral index $\propto E^{-2}$). The final flux upper limits at 95\% confidence level are:
\begin{equation}
F_\gamma^{95\%}  (E_\gamma > 10, 20, 40~{\rm EeV}) < (1.9,~1.0,~0.49) \times 10^{-3}~{\rm km^{-2}~yr^{-1}~sr^{-1}}~~.
\end{equation}
As shown in Figure \ref{fig.Diffuse} (right) these limits are the most stringent above 10~EeV and non-acceleration models are severely disfavored by these results. Furthermore, optimistic (proton dominated) GZK scenarios are in reach, especially taking the advanced photon/hadron discrimination of the Auger upgrade AugerPrime into account, cf.\ last section. In addition to the expected GZK photon flux prediction of a pure proton and iron composition, derived in \cite{SarkarICRC}, two mixed composition scenarios based on the results given in \cite{MatteoICRC} are shown in Figure \ref{fig.Diffuse} (right) using CRPropa 3 \cite{1475-7516-2016-05-038} as cosmic ray propagation tool \cite{DavidPrivate}. The optimistic (high photon) scenario refers to the second minimum in the spectral index / maximum rigidity fit in \cite{MatteoICRC} which is more in line with standard models of cosmic ray acceleration. The best fit to Auger data, however, indicates to a (low photon) scenario where the flux is mostly limited by the maximum energy at the sources resulting in a hard spectral index ($\lesssim 1$) and a low rigidity dependent cutoff ($\log(R_{\rm cut}/{\rm V}) \lesssim 19$) where only a small photon flux at UHE is expected, cf.\ Figure \ref{fig.Diffuse} (right) lower GZK mixed line. Note, that only statistical uncertainties are shown and no systematics on the photon background and source evolution are included. As a result of the reduced maximum energy at the source, expected photon fluxes in the mixed scenario are lower than in \cite{SarkarICRC} where the maximum energy is fixed to $Z \times 10^{21}$~eV.

\section{DIRECTIONAL SEARCH FOR PHOTON POINT SOURCES}
The search for photon point sources includes the directional information of each event. The flux of photons from a single direction can be detected by an excess of air showers arriving from that direction within the angular resolution of the experiment. While charged particles are deflected in galactic and extragalactic magnetic fields, neutral particles point back to their production site. The search for neutron point sources with the Pierre Auger Observatory has been published in \cite{NeutronBlind,TargetedNeutron} and in the following the directional search for photon point sources is outlined as discussed in more detail in \cite{BlindPhoton}. The basic idea is to reduce the background contamination by selecting only events similar to simulated photon air showers. The search is limited to the energy range between $10^{17.3}$~eV and $10^{18.5}$~eV using hybrid events, i.e.\ events using information from the FD and SD detector, in an declination range between $-85^\circ$ and $+20^\circ$. In this field of view, 526200 target directions are equally distributed with an average target separation of $\sim 0.3^\circ$. A top-hat search window with radius $1^\circ$ centered on each target direction is used, a compromise between possible non-Gaussian shape of the error distribution, low event statistics, and the angular resolution of 0.7$^\circ$. To ensure good energy and directional reconstruction the search is limited to zenith angles smaller than 60$^\circ$ with good reconstruction of the shower geometry. In addition, several quality requirements ensure a reliable profile reconstruction. More details are given in \cite{BlindPhoton}. Finally, the total event sample consists of 241466 events. 

To reduce the isotropic hadron background contamination several mass-sensitive observables are defined and combined using the multivariate analysis technique of boosted decision trees \cite{cart84,Schapire1990}.  The three FD observables are the depth of shower maximum $X_{\rm max}$, the ratio of the reconstructed energy using a Greisen fit to the profile and the standard Gaisser-Hillas fit function, and the corresponding reduced $\chi^2_{\rm Gr} / {\rm ndof}$ of the fit. The Greisen function \cite{Greisen1956} was derived from purely electromagnetic cascade theory and is expected to describe photon air showers better compared to hadron induced showers. Two additional SD observables are introduced. The $S_b$-parameter is sensitive to different lateral distribution functions taking into account the signal strength at various distances from the shower core \cite{Ros2011140}. Finally, the shape parameter, defined as the distance and inclination angle corrected ratio between early and late arriving integrated signal, is sensitive to the spread in arrival times of shower particles (a larger spread is expected in deep developing showers). More details on the observables can be found in \cite{BlindPhoton}. The multivariate response function, $\beta$, is trained with Monte Carlo proton and photon simulations. 

\begin{figure}[t!]
  \centerline{\includegraphics[width=480pt]{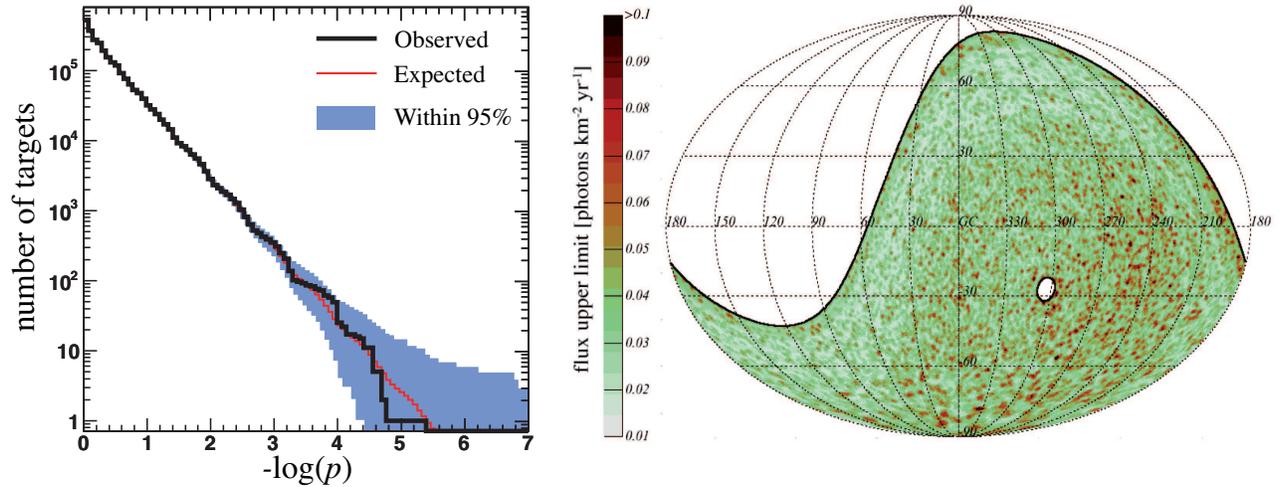}}
  \caption{\textit{Left panel:} Distribution of $p$ values. The thick black line illustrates the observed and the red line the expected distribution. The blue shaded area denotes the 95\% containment of simulated datasets.  \textit{Right panel:} Celestial map of photon flux limits from point sources using galactic coordinates. The declination limit of $-85^\circ$ and $+20^\circ$ is denoted by a think black line.}
  \label{fig.Directional}
\end{figure}

The impact of an isotropic background contribution is estimated using the scrambling technique \cite{CASSIDAY1990291} by connecting randomly arrival directions (in local coordinates) with a Coordinated Universal Time from a pool of measured events and repeating this procedure 5000 times. The expected directional photon exposure is derived from time-dependent detector simulations described in \cite{ExposureAuger}. 

For each target direction only a subset of the recorded data, selected as ``photon-like'' according to the $\beta$ distribution, is used. This optimized $\beta_{\rm cut}$ is a function of the expected number of events of a target direction and determined by minimizing the expected upper limit, using the definition of Zech in \cite{ZECH1989608}, assuming that the expected number of events $b$ is equal to the measured number of events $n$. After applying this optimized $\beta_{\rm cut}$ the probability $p$ of obtaining a test statistic at least as extreme as the one that was actually observed is calculated
\begin{equation}
p = {\rm Poiss} (\geq n_{\rm data}^\beta | n_b^\beta )~,
\end{equation}
where $n_{\rm data}^\beta$ is the observed number and $n_b^\beta$ the expected number of events. The subscript $\beta$ indicates event numbers after applying $\beta_{\rm cut}$. The distribution of $p$ values for all target directions is shown in Figure \ref{fig.Directional} (left). The tail of the distribution is in agreement with isotropy and the minimum $p$-value observed is $p_{\rm min}=4.5\times10^{-6}$ corresponding to a chance probability that $p_{\rm min}$ is observed
anywhere in the sky of $p_{\rm chance} = 36\%$.

Photon flux upper limits from point sources at 95\% confidence level are calculated as 
\begin{equation}
f^{\rm UL} = \frac{n_s^{\rm Zech}}{n_{\rm inc} \cdot \mathcal{E}_{\rm{\beta}}}~,
\end{equation}
where $n_s^{\rm Zech}$ is the upper limit on the number of photons obtained using Zech's method \cite{ZECH1989608,BlindPhoton}, $\mathcal{E}_{\rm{\beta}}$ the directional photon exposure and  $n_{\rm inc} = 0.9$ a correction factor to take into account signal events outside the top-hat search window. A celestial map of upper limits is shown in Figure \ref{fig.Directional} (right). The particle mean upper limit is 0.035~km$^{-2}$~yr$^{-1}$, with a maximum of 0.14~km$^{-2}$~yr$^{-1}$ corresponding to to an energy flux limit of 0.06~eV~cm$^{-2}$~s$^{-1}$ and 0.25~eV~cm$^{-2}$~s$^{-1}$,
respectively, assuming an $E^{-2}$ energy spectrum.

These limits are of astrophysically interest in all parts of the exposed sky. Measured gamma ray fluxes exceed 1~eV~cm$^{-2}$~s$^{-1}$ in the TeV energy range for some sources with a spectral index $\propto E^{-2}$~\citep{Hinton2009}. Extrapolating this flux to EeV energies one expects 1~eV~cm$^{-2}$~s$^{-1}$ in the EeV range. No energy flux that strong in EeV photons is observed from any target direction and the continuation can be excluded with more than $5 \sigma$. Furthermore, H.E.S.S.\ published the spectrum from the galactic center region without an indication of a cutoff or a spectral break implying that our Galaxy hosts petaelectronvolt accelerators ``PeVatrons'' \citep{HESSPevatron}. The naive extrapolation of this particle flux to the energy band of this paper gives an integrated flux expectation of $J_{\rm int}^{\rm exp} = 0.045$~km$^{-2}$~yr$^{-1}$. Comparing this value with the mean upper limit of 0.035~km$^{-2}$~yr$^{-1}$, derived from Auger observations in the EeV range, already indicate that current photon flux limits from the galactic center may constrain a continuation to ultra-high energies and set upper bounds on the cutoff energy. A more detailed analysis on photon flux upper limits of specific targets, including the galactic center, is currently in preparation.

\section{CONCLUSION AND OUTLOOK}
Despite a sensitive search for UHE photons, no positive detection has been found with the Pierre Auger Observatory or any other experiment. Current fractional and flux upper limits presented in this paper  severely constrain non-acceleration models and start to touch optimistic predictions of GZK-type production processes. Directional searches for photon point sources did not find any indiction of nearby photon emitters in the EeV energy range. Corresponding upper limits of photon point sources could be set over a wide declination range and are of astrophysically interest, e.g.\ when extrapolating measured TeV photon fluxes to UHE. 

With new triggering algorithm and additional composition sensitivity of the planned upgrade of the Pierre Auger Observatory AugerPrime will increase the sensitivity to photons significantly \cite{Aab:2016vlz}. By 2024 it is expected to lower the photon limits to reach even conservative predictions of GZK photons. The directional search will also profit from the upgrade enabling sophisticated point source studies or even allow a detection of photons from specific sources.



\nocite{*}
\bibliographystyle{aipnum-cp}%
\bibliography{myBib}%

\end{document}